\documentclass[acmtoms]{acmtrans2m}

\acmVolume{V}
\acmNumber{N}
\acmYear{YY}
\acmMonth{Month}

\usepackage{amsmath}
\usepackage{color}
\usepackage{graphicx}
\usepackage{caption}
\usepackage{subcaption}
\usepackage{booktabs}
\usepackage{colortbl}
\usepackage{enumerate}

\usepackage[dotinlabels]{titletoc}
\usepackage{array}

\newcommand{\algname}{\textcolor{white}{ (NAME)}}

\newcommand{\genabel}{{GenABEL}}
\newcommand{\fastlmm}{FaST-LMM}
\newcommand{\gw}{{\sc gwfgls}}
\newcommand{\flmm}{{\sc fast-lmm}}

\newcommand{\eig}{{\sc mt-gwas}}
\newcommand{\eigone}{{\sc mt-gwas-ct}}
\newcommand{\eigtwo}{{\sc mt-gwas-it}}

\newcommand{\bi}{\begin{itemize}}
\newcommand{\ei}{\end{itemize}}

\usepackage{listings}
\usepackage{caption}
\DeclareCaptionFont{white}{\color{white}}
\DeclareCaptionFont{black}{\color{black}}
\DeclareCaptionFormat{listing}{\colorbox[cmyk]{0.23, 0.15, 0.15,0.01}{\parbox{\textwidth}{\centering #1\algname{}#2#3}}}
\title{
  Computing Petaflops over Terabytes of Data:\\ 
  The Case of Genome-Wide Association Studies
}
            
\author{
DIEGO FABREGAT-TRAVER \\
Aachen Institute for Advanced Study in Computational Engineering Science, \\RWTH Aachen
\and
PAOLO BIENTINESI \\
Aachen Institute for Advanced Study in Computational Engineering Science, \\RWTH Aachen
}
            
\begin{abstract} 
In many scientific and engineering applications, one has to solve 
not one but a sequence of instances of the same problem.
Often times, the problems in the sequence are linked in a way that allows
intermediate results to be reused.
A characteristic example for this class of applications is given by
the Genome-Wide Association Studies (GWAS), 
a widely spread tool in computational biology.
GWAS entails the solution of up to trillions ($10^{12}$) of 
correlated generalized least-squares problems, posing a daunting
challenge: the performance of petaflops 
($10^{15}$ floating-point operations) over terabytes of data.

In this paper, we design an algorithm
for performing GWAS on multi-core architectures.
This is accomplished in three steps.
First, we show how to exploit the relation among successive problems,
thus reducing the overall computational complexity.
Then, through an analysis of the required data transfers, we identify 
how to eliminate any overhead due to input/output operations.
Finally, we study how to decompose computation into tasks to be 
distributed among the available cores, 
to attain high performance and scalability.

We believe the paper contributes valuable guidelines 
of general applicability for computational scientists on how to 
develop and optimize numerical algorithms. 
\end{abstract}
            

            
\begin{document}
            
\begin{bottomstuff} 
  Authors' addresses:
  Diego Fabregat, AICES, RWTH Aachen, Aachen, Germany, 
  {\tt fabregat@aices.rwth-aachen.de}. 
  \newline
  Paolo Bientinesi, AICES, RWTH Aachen, Aachen, Germany, 
  {\tt pauldj@aices.rwth-aachen.de}. 
\end{bottomstuff}
            
\maketitle

\section{Introduction}
\label{sec:intro}
Many traditional linear algebra libraries, such as LAPACK~\cite{laug}
and ScaLAPACK~\cite{slug}, and tools such as Matlab, focus on
providing efficient building blocks for solving one instance of many
standard problems.  By contrast, engineering and scientific
applications often originate multiple instances of the same problem,
thus leading to a ---possibly very large--- sequence of independent
library invocations.  The drawback of this situation is the missed
opportunity for optimizations and data reuse; in fact, due to their
black-box nature, libraries cannot avoid redundant computations or
exploit problem-specific properties across invocations.  The
underlying theme of this paper is that a solution scheme devised for a
specific sequence may be much more efficient than the repeated
execution of the best single-instance
routine~\cite{SDP-UHP,Di_Napoli2012:160}.

Genome-wide Association Studies (GWAS) clearly exemplify this issue,
posing a formidable computational challenge: the solution of billions,
or even trillions, of correlated generalized least-squares (GLS)
problems.  As described below, a solution based on the traditional
black box approach is entirely unfeasible.  With this paper we
demonstrate that by tackling all the problems as a whole, and by
exploiting application specific knowledge in combination with a
careful parallelization scheme, it becomes possible for biologists to
complete GWAS in matter of hours.

Any algorithm to perform GWAS needs to address the following issues. 
\begin{itemize}
\item 
  {\bf Complexity.}
  In a representative study, one has to solve a grid 
  of $m \times t$ generalized least-squares problems, where
  $m$ and $t$ are of the order of millions and hundreds of thousands, respectively.
  The complexity for the solution of one problem (of size $n$) in isolation is $O(n^3)$
  floating point operations (flops), adding up to $O(m t n^3)$ flops for the entire study.
  In the case of the largest problem addressed in our experiments 
  (Section~\ref{sec:experiments}; $m = 10^6$, $t=10^5$, and $n=10^3$), 
  this approach would require the execution of roughly $10^{23}$ flops; 
  even if performed at the theoretical peak of {\it Sequoia}, the fastest supercomputer in the 
  world,\footnote{\url{http://www.top500.org/}, as of October 2012.} 
  the computation would take more than 2 months.
  Here we illustrate how, by exploiting the structure that links different problems, 
  the complexity reduces to $O(m t n)$, and the same experiment
  on a 40-core node completes in about 12 hours.

\item 
  {\bf Size of the datasets.}
  GWAS entails the processing of terabytes of data. Specifically,
  the aforementioned analysis involves reading 10 GBs and writing 3.2 TBs.
  Largely exceeding even the combined main memory of current typical clusters,
  these requirements demand an out-of-core mechanism to efficiently handle datasets residing on disk.
  In such a scenario, data is partitioned in {\em slabs}, and it becomes critical 
  to determine the most appropriate traversal,
  as well as the size and shape of each slab.
  The problem is far from trivial, because these factors affect 
  the amount of transfers and computation performed,
  as well as the necessary extra storage.
  By modeling all these factors, we find the best traversal direction, 
  and show how to determine the shape and size of the slabs to achieve a complete overlap  
  of transfers with computations. As a result, irrespective of the data size, the efficiency 
  of our in-core solver is sustained.
\item 
  {\bf Parallelism.}
  The partitioning of data in slabs that fit in main memory translates 
  to the computation of the two-dimensional grid of GLSs 
  in terms of sub-grids or {\em tiles} of such problems.
  The computation of each tile
  must be organized to exploit multi-threaded parallelism and to attain
  scalability even with a large number of cores.
  To this end, we present a study on how to 
  decompose the tiles into smaller computational tasks
  ---to take advantage of the cache memories--- 
  and how to distribute such tasks among the computing cores.
\end{itemize}

Our study of these three issues is not specifically tied to GWAS; we
keep the discussion in general terms, to concentrate on the methodology
rather than on the problem at hand.  We believe this paper contributes
valuable guidelines for computational scientists on how to optimize
numerical algorithms.
For the specific case of GWAS, 
thanks to the combination of a lower complexity, a perfect overlapping
of data movement with computation, and a nearly perfect scalability,
we outperform the current state-of-the-art tools, \genabel{} and
\fastlmm{}~\cite{genabel,Lippert2011}, 
by a factor of more than 1000.

Section~\ref{sec:problem} introduces GWAS both in biology and linear algebra terms. 
In Section~\ref{sec:core-algorithm}, we detail how our algorithm for GWAS exploits 
application-specific knowledge to reduce the asymptotical cost of the best existing 
algorithms, while in Section~\ref{sec:ooc} we analyze the required data transfers, 
and discuss the application of out-of-core techniques to completely hide the 
overhead due to input/output operations.
In Section~\ref{sec:parallelism}, we tailor our algorithm to exploit shared-memory
parallelism, and we provide performance results in Section~\ref{sec:experiments}.
Future work is discussed in Section~\ref{sec:futurework}, 
and conclusions are drawn in Section~\ref{sec:conclusions}.

\section{Multi-trait Genome-Wide Association Studies}
\label{sec:problem}

In a living being, 
observable characteristics ---{\it traits} or {\it phenotypes}--- 
such as eye color, height, and susceptibility to disease, are influenced 
by information encoded in the genome.
The identification of specific regions of the genome 
---{\em single-nucleotide polymorphisms} or {\it SNPs}---
associated to a given trait
enhances the understanding of the trait, and, in the case of diseases, 
facilitates prevention and treatment. 
Genome-wide Association Studies (GWAS) 
are a powerful statistical tool for locating the SNPs
involved in the control of a trait~\cite{10.1371/journal.pgen.1001256-short,Levy2009-short,Speliotes2010-short}. 
The simultaneous analysis of many phenotypes is the objective of the so called multi-trait GWAS.

Every year, computational biologists publish hundreds of GWAS-related
papers~\cite{GWAScatalog}, with a clear trend towards analyses that
include more and more SNPs.  Ideally, the computational biologists aim
at testing the whole human genome against as many traits as possible.
Mathematically, for each SNP $X_i$ and trait $y_j$ 
( $i \in [1\dots m], j \in [1\dots t]$), 
one has to solve the generalized least-squares problem (GLS)
\begin{equation}
\label{eq:probDef}
b_{ij} := (X_i^T M_j^{-1} X_i)^{-1} X_i^T M_j^{-1} y_j 
\end{equation}
where
\begin{itemize}
\item 
$y_j$ is the vector of observations; 
\item 
$X_i$ is the design matrix; 
\item 
$M_j$ is the covariance matrix, representing dependencies among observations; and
\item 
the vector $b_{ij}$ expresses the relation between a variation in the SNP ($X_i$)
and a variation in the trait ($y_j$).
\end{itemize}
In multi-trait GWAS, the covariance matrix satisfies
\begin{equation} 
  \label{eq:m}
  M_j    := \sigma^2_j \cdot (h^2_j \Phi + (1 - h^2_j) I),
\end{equation}
where
\begin{itemize}
\item 
the {\it kinship} matrix $\Phi$ contains the relationship among all studied individuals,
\item 
$\sigma^2$ and $h^2_j$ are trait-dependent scalar estimates, and 
\item 
$I$ is the identity matrix.
\end{itemize}
The number of SNPs, $m$, ranges between $10^6$ and $10^8$ 
(180.000.000 for the full human genome), and
the number of traits, $t$, is either 1 (single-trait analysis), 
or ranges between $10^4$ and $10^5$.

\begin{figure}
    \centering
    \includegraphics[scale=.58]{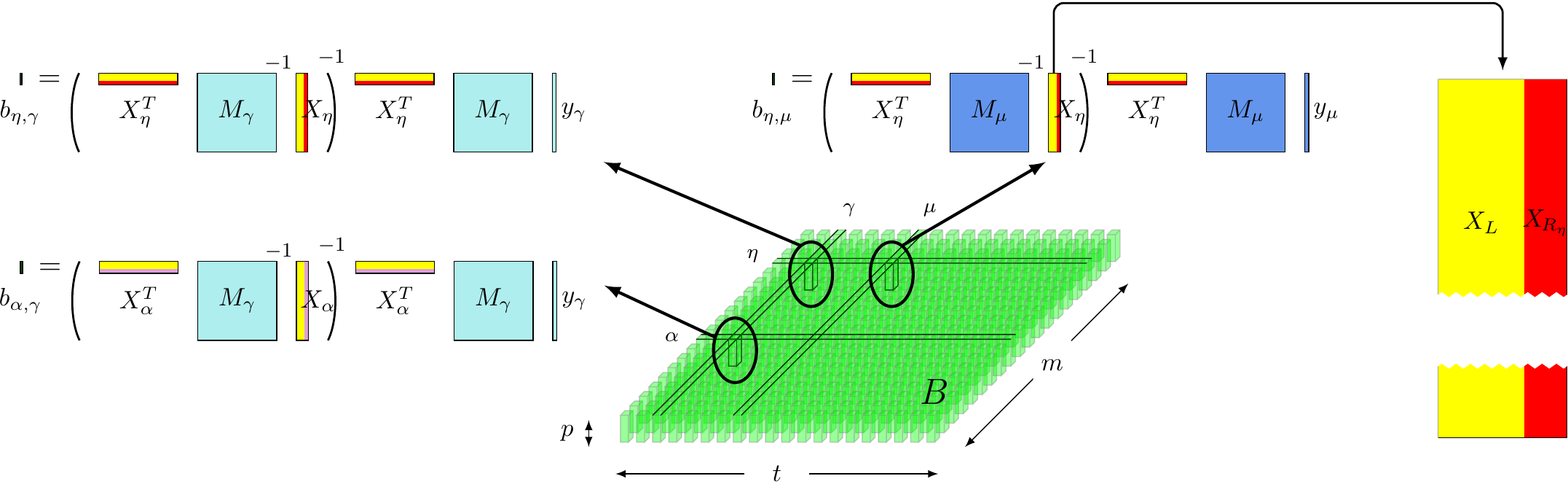}
\caption{Interpretation of GWAS as a two-dimensional sequence 
of generalized least-squares 
problems ($ b := (X^T { M }^{-1} X)^{-1} X^T { M }^{-1} y$).
GWAS with multiple phenotypes requires the solution of $m \times t$ correlated
GLS problems, originating a three-dimensional object $B$ of size 
$m \times t \times p$.  
As suggested by the colors,
along the $t$ direction the covariance matrix $M$ and the phenotype
$y$ vary, while the design matrix $X$ does not; conversely, in the $m$
direction, $M$ and $y$ are fixed, while $X$ varies.
The colors also hint at the fact that 
$X$ can be viewed as consisting of two parts, $X_L$ and
$X_R$, where the former is constant across the entire grid and the
latter changes along $m$.
}
\label{fig:ProblemDescription}
\end{figure}

In Fig.~\ref{fig:ProblemDescription}, 
we provide a visual interpretation of the problem: 
Each point in the grid, i.e., the vertical vector of size $p$ at position $(i,j)$,
corresponds to the solution of one GLS
(association between the $i$-th SNP and the $j$-th trait).
A column in the figure represents a single-trait analysis 
(association between all the SNPs and a given trait);
in this case, $y_j$ as well as $h^2_j$ and $\sigma^2_j$, 
and therefore $M_j$, are fixed.
The full grid depicts the whole multi-trait analysis;
as expected, $y$, $h^2$, $\sigma^2$, and $M$ vary along the $t$ dimension, 
while the SNP $X_i$ is fixed; the kinship matrix $\Phi$ is constant 
throughout the two-dimensional grid.

In linear algebra terms, 
Eq.~\eqref{eq:probDef} solves a linear regression with non-independent outcomes, 
where
$b_{ij} \in \mathcal{R}^{p}$, 
$X_i \in \mathcal{R}^{n \times p}$ is full rank, 
$M_j \in \mathcal{R}^{n \times n}$ is symmetric positive definite (SPD),
$y_j \in \mathcal{R}^{n}$, 
$\Phi \in \mathcal{R}^{n \times n}$ is symmetric,
$I \in \mathcal{R}^{n \times n}$, and
$\sigma^2_j$ and $h^2_j \in \mathcal{R}$.
Moreover, 
the design matrix $X_i$ presents a special structure:  
each $X_i$ may be partitioned as $(X_L \; | \; X_{R_i})$, 
where $X_L \in R^{n \times (p-1)}$ is the same for all $X_i$'s,
while $X_{R_i} \in R^{n \times 1}$ varies.
The sizes are 
$10^3 \le n \le 10^4$ and 
$2 \le p \le 20$.
The quantities $X_i$, $\Phi$, $h^2_j$, $\sigma^2_j$, and $y_j$ are known, and the vector $b_{ij}$ is to be computed.

\subsection{Related Work}

The black-box nature of traditional numerical libraries limits them to
offering routines for the zero-dimensional case, a point in the grid, and use 
such kernel repeatedly for each problem in the sequence. As discussed previously,
such an approach is absolutely unviable.
Instead, GWAS-specific tools, such as \genabel{} and \fastlmm{}, 
incorporate and exploit knowledge specific to the application.
However, these tools still present limitations in all three of the
GWAS challenges: 1) they are tailored for a one-dimensional sequence of problems,
individual columns of Fig.~\ref{fig:ProblemDescription}; for the solution
of the entire two-dimensional grid they only offer the possibility of 
repeatedly using
their one-dimensional solver.
From a multi-trait GWAS perspective, this still represents a black box
approach, missing opportunities for further computation reuse;
2) they incorporate rudimentary out-of-core mechanisms that lack 
the overlapping of data transfers with computation, 
thus incurring a considerable overhead;
and 3) they attain only poor scalability.

\subsection{Terminology}

Here we give a brief description of the acronyms used throughout the paper.
\begin{itemize}
\item GWAS: Genome-Wide Association Studies.
\item GLS: Generalized Least-Squares problems.
\item \genabel{}: A framework for statistical genomics, 
  including one of the most widely used 
  packages to perform {\sc gwas}.
\item \gw{}: GenABEL's state-of-the-art routine for GWAS.
\item \flmm{}: The most recent high-performance tool for GWAS.
\item \eig{}: our novel solver for multi-trait GWAS.
\end{itemize}
Additionally, we will reference many times
\begin{itemize}
\item BLAS (Basic Linear Algebra Subprograms)~\cite{BLAS3}, and
\item LAPACK (Linear Algebra PACKage),
\end{itemize}
as the de-facto standard libraries 
for high-performance dense linear algebra computations.

\section{The multi-trait GWAS algorithm}
\label{sec:core-algorithm}

\eig{}, a novel algorithm for multi-trait GWAS is introduced: 
First we describe a simplified version that solves one single GLS instance,
and then show how the algorithm can be tailored 
for Eqs.~\eqref{eq:probDef} and \eqref{eq:m}, 
the solution of the whole GWAS.
We stress that while \eig{} is suboptimal for the solution of one single GLS, 
it exploits the specific structure and properties of the full problem,  
and dramatically reduces its computational complexity, 
thus achieving remarkable speedups.

\subsection{Single-instance}

The fastest approach for Eq.~\eqref{eq:probDef} involves computing 
the Cholesky factorization of the matrix $M$~\cite{Paige-FastGLS-1979}, for a cost of $\frac{n^3}{3}$ flops.
In the context of GWAS, in which $M$ varies with the index $j$ (Eq.~\eqref{eq:m}), 
such a factorization has to be performed $t = 10^4$--$10^5$ times, 
thus originating a computational bottleneck.
Instead, the main idea behind \eig{} is to take advantage 
of the fact that $M$ results from scaling and shifting the matrix $\Phi$, which 
remains constant for all $i$'s and $j$'s. Then, computing the 
eigendecomposition $Z \Lambda Z^T = \Phi$ ($Z$ orthogonal, $\Lambda$ diagonal),
\begin{align*}
M_j :&= \sigma^2_j ( h^2_j Z \Lambda Z^T + (1-h^2_j) I) \\
    &= Z D_j Z^T, \ \ \rm{where} \ \ D_j = \sigma^2_j ( h^2_j \Lambda + (1-h^2_j) I).
\end{align*}
While the eigendecomposition is much more expensive than the Cholesky factorization,
$\frac{10}{3} n^3$ vs. $\frac{n^3}{3}$ flops, 
it allows us to express $M_j^{-1}$, {for any} $j$, with only $O(n)$ 
additional flops:
\[
M_j^{-1} = Z D_j^{-1} Z^T.
\]

After this initial step, Eq.~\eqref{eq:probDef} can be rewritten as
\[
  b_{ij} := (X_i^T Z D_j^{-1} Z^T X_i)^{-1} X_i^T Z D_j^{-1} Z^T y_j.
\]
Since $M_j$ is SPD, its eigenvalues $D_j$ are all positive. 
Accordingly, the algorithm may take advantage of the symmetry of 
the expression $X_i^T Z D_j Z^T X_i$, and save computations. 
This is accomplished by computing the inverse of the square root 
of each entry of $D_j$ ($K_j := D^{-\frac{1}{2}}_j)$, resulting in 
$$ b_{ij} := (X_i^T Z K_j K^T_j Z^T X_i)^{-1} X_i^T Z K_j K^T_j Z^T y_j. $$
Next, the products $X'_i := Z^T X_i$ and $y'_j := Z^T y_j$ are computed
$$ b_{ij} := (X'^T_i K_j K^T X'_i)^{-1} X'^T_i K_j K^T y'_j, $$
and also $W_{ij} := K^T_j X'_i$ and $ v_{j} := K^T_j y'_j$
$$ b_{ij} := (W^T_{ij} W_{ij})^{-1} W^T_{ij} v_{j}. $$

What remains is an ordinary least-squares problem. Numerical
considerations allow us to safely rely on forming
$S_{ij} := W^T_{ij} W_{ij}$ without incurring instabilities. 
The algorithm completes by computing $b_{ij} := W^T_{ij} v_{j}$ and
solving the corresponding SPD linear system $b_{ij} := S_{ij}^{-1} b_{ij}$. 
\eig{} is detailed in Algorithm~\ref{alg:singleGLS}. 

\begin{center}
\renewcommand{\lstlistingname}{Algorithm}
\renewcommand{\algname}{}
\begin{minipage}{0.60\linewidth}
\begin{lstlisting}[caption=\eig{} for the solution of a single instance of the generalized
least-squares problem., label=alg:singleGLS, escapechar=!]
  $Z \Lambda Z^T = \Phi$
  $D_j := \sigma^2_j (h^2_j \Lambda + (1 - h^2_j) I)$
  $K_j K_j^T = D_j^{-\frac{1}{2}}$
  $X'_i := Z^T X_i$
  $y'_j := Z^T y_j$
  $W_{ij} := K_j^T X'_i$
  $v_j := K_j^T y'_j$
  $S_{ij} := W_{ij}^T W_{ij}$
  $b_{ij} := W_{ij}^T v_j$
  $b_{ij} := S_{ij}^{-1} b_{ij}$
\end{lstlisting}
\end{minipage}
\end{center}

\subsection{Tailoring for the two-dimensional sequence}
\label{sec:tailoringtwod}

We now discuss how to extend Algorithm~\ref{alg:singleGLS}
for the solution of the two-dimensional sequence of GLSs specific to GWAS.
The general objective is to identify opportunities for reusing partial calculations 
across different problems, in order to reduce the overall cost.

As a first step, we expose further application-specific knowledge: 
between any two matrices $X_{i_1}$ and $X_{i_2}$, only the right portion changes.
In Algorithm~\ref{alg:Xexposed}, every appearance of $X_i$ is 
replaced with its partitioned counterpart $( X_L \;|\; X_{R_i} )$, and 
the partitioning is propagated.
The subscripts $L$, $R$, $T$, and $B$ stand for $L$eft, $R$ight, $T$op, and $B$ottom,
respectively. Since $S$ is symmetric, the star in the top-right quadrant 
indicates the transpose of $S_{BL_{ij}}$, and this quadrant
does not need to be either stored or computed.
This intricate refinement of the algorithm is 
absolutely necessary to expose for each operand, 
which portion varies along which dimension.

\begin{center}
\renewcommand{\lstlistingname}{Algorithm}
\renewcommand{\algname}{}
\begin{minipage}{0.60\linewidth}
\renewcommand{\arraystretch}{1.2}
\large
\begin{lstlisting}[caption=Single-instance version of \eig{}; the structure of $X_i$ is exposed., label=alg:Xexposed, escapechar=!]
  $Z \Lambda Z^T = \Phi$
  $D_j := \sigma^2_j (h^2_j \Lambda + (1 - h^2_j) I)$
  $K_j K_j^T = D_j^{-\frac{1}{2}}$
  $(X'_L \;|\; X'_{R_i}) := Z^T (X_L \;|\; X_{R_i})$
  $y_j' := Z^T y_j$
  $(W_{L_j} \;|\; W_{R_{ij}}) := K_j^T (X'_L \;|\; X'_{R_i})$
  $v_j := K_j^T y_j'$
  $\left( 
  \begin{array}{c | c} 
    S_{TL_j} & \star \\\hline 
    S_{BL_{ij}} & S_{BR_{ij}}
  \end{array} 
  \right) := 
  \left( 
  \begin{array}{c | c} 
    W_{L_j}^T W_{L_j} & \star \\\hline 
    W_{R_{ij}}^T W_L & W_{R_{ij}}^T W_{R_{ij}}
  \end{array} 
  \right)$
  $\left( 
    \begin{array}{c} 
		b_{T_j} \\\hline 
		b_{B_{ij}} 
    \end{array} 
  \right) :=
  \left( 
    \begin{array}{c} 
	   W_{L_j}^T v_j \\\hline 
	   W_{R_{ij}}^T v_j 
    \end{array} 
  \right)$
   $b_{ij} := S_{ij}^{-1} b_{ij}$
\end{lstlisting}
\end{minipage}
\end{center}

Next, we wrap Algorithm~\ref{alg:Xexposed} 
with a double loop, corresponding to the traversal
of the two-dimensional $m$-$t$ grid, 
and reorganize the operations, 
aiming at eliminating redundant computation.
Both traversals of the  grid 
---by rows ({\tt for i, for j}) and by columns ({\tt for j, for i})--- 
are provided, in Algorithms~\ref{alg:twoDij} and~\ref{alg:twoDji}, respectively.
We will use the latter to describe how the computation can be rearranged; 
the same reasoning applies to the former.

For each operation in the algorithm, 
the dependencies on the indices $i$ and $j$ 
are determined by the left-hand side operand(s).
Any operation whose left-hand side does not include any subscript 
is invariant across the two-dimensional sequence; therefore it can be 
computed once and reused in every other iteration of the loops.
This is the case for the eigendecomposition of $\Phi$, and the computation of $X'_L$. 
Operations that only vary across the $t$ dimension (subscript $j$), are performed
once per iteration over $t$ ---the outer loop--- 
and reused across the iterations over $m$ ---the inner loop---.
Finally, the operations labeled with $i$ or $i,j$ are placed in the innermost loop.

\begin{center}
\renewcommand{\lstlistingname}{Algorithm}
\renewcommand{\algname}{}
\begin{minipage}{0.45\linewidth}
\renewcommand{\arraystretch}{1.2}
\large
\begin{lstlisting}[caption=Solution of the two-dimensional grid of GLS problems 
depicted in Fig.~\ref{fig:ProblemDescription}. Traversal by rows.,label=alg:twoDij, escapechar=!]
 $Z \Lambda Z^T = \Phi$
 $X'_L := Z^T X_L$
 for 1 $\le$ i $\le$ m
   $X'_{R_i} := Z^T X_{R_i}$
   for 1 $\le$ j $\le$ t
     $D_j := \sigma^2_j (h^2_j \Lambda + (1 - h^2_j) I)$
     $K_j K_j^T = D_j^{-\frac{1}{2}}$
     $y_j' := Z^T y_j$
     $W_{L_j} := K_j^T X'_L$
     $W_{R_{ij}} := K_j^T X'_{R_i}$
     $v_j := K_j^T y_j'$
     $S_{TL_j} := W_{L_j}^T W_{L_j}$
     $S_{BL_{ij}} := W_{R_{ij}}^T W_{L_j}$
     $S_{BR_{ij}} := W_{R_{ij}}^T W_{R_{ij}}$
     $b_{T_j} := W_{L_j}^T v_j$
     $b_{B_{ij}} := W_{R_{ij}}^T v_j$
     $b_{ij} := S_{ij}^{-1} b_{ij}$
\end{lstlisting}
\end{minipage}
\hfill
\renewcommand{\lstlistingname}{Algorithm}
\renewcommand{\algname}{}
\begin{minipage}{0.46\linewidth}
\renewcommand{\arraystretch}{1.2}
\large
\begin{lstlisting}[caption=Solution of the two-dimensional grid of GLS problems 
depicted in Fig.~\ref{fig:ProblemDescription}. Traversal by columns.,label=alg:twoDji, escapechar=!]
 $Z \Lambda Z^T = \Phi$
 $X'_L := Z^T X_L$
 for 1 $\le$ j $\le$ t
   $D_j := \sigma^2_j (h^2_j \Lambda + (1 - h^2_j) I)$
   $K_j K_j^T = D_j^{-\frac{1}{2}}$
   $y_j' := Z^T y_j$
   $W_{L_j} := K_j^T X'_L$
   $v_j := K_j^T y_j'$
   $S_{TL_j} := W_{L_j}^T W_{L_j}$
   $b_{T_j} := W_{L_j}^T v_j$
   for 1 $\le$ i $\le$ m
     $X'_{R_i} := Z^T X_{R_i}$
     $W_{R_{ij}} := K_j^T X'_{R_i}$
     $S_{BL_{ij}} := W_{R_{ij}}^T W_{L_j}$
     $S_{BR_{ij}} := W_{R_{ij}}^T W_{R_{ij}}$
     $b_{B_{ij}} := W_{R_{ij}}^T v_j$
     $b_{ij} := S_{ij}^{-1} b_{ij}$
\end{lstlisting}
\end{minipage}
\end{center}

As the reader might have noticed, the algorithm still performs
redundant computations: Lines 6--9, 11--12, and 15 in
Algorithm~\ref{alg:twoDij} and line 12 in Algorithm~\ref{alg:twoDji}
depend only on the dimension traversed by the inner loop, and 
are therefore recomputed at each iteration of the outer loop.
This flaw can be resolved by precomputing the quantities outside
of the double loop, and then accessing them from within the
loops. While mathematically and algorithmically possible, the approach
poses practical constraints on Algorithm~\ref{alg:twoDij}, as it would
require a fairly large amount of extra storage.
The solution is instead applicable in Algorithm~\ref{alg:twoDji}: 
The operation $X'_{R_i} := Z^T X_{R_i}$ may overwrite $X_{R_i}$, making 
the size of temporary storage negligible.
Let us stress the significance of this improvement: 
by avoiding redundant calculations, 
Algorithm~\ref{alg:twoDji} saves $2 t m n^2$ flops,
thus drastically lowering the cost with respect to the other 
state-of-the-art algorithms, and 
eliminates the main computation bottleneck.
Henceforth, Algorithm~\ref{alg:twoDji} is our algorithm of choice.

Practical considerations allow us for one more optimization. 
Notice that operation $X'_{R_i} := Z^T X_{R_i}$ is performed independently
for each $X_{R_i}$, resulting in $m$ matrix-vector multiplications. 
Instead, these matrix-vector operations may be bundled
together in a single large matrix-matrix product, known to be a much
more efficient operation.
Similar reasoning applies to the operation at line 6, $y'_j := Z^T y_j$. 

\eig{}, the final algorithm that includes all these optimizations,  
is provided in Algorithm~\ref{alg:core}. There, $\mathcal{X}_R$ and 
$\mathcal{Y}$ are used to represent the collection of all 
$X_R$'s and all $y$'s, respectively:
$\mathcal{X}_R = (X_{R_1} \; | \; X_{R_2} \; | \; \ldots \; | \; X_{R_m})$ and 
$\mathcal{Y} = (y_1 \; | \; y_2 \; | \; \ldots \; | \; y_t)$.
The second and third columns indicate, respectively, for each individual operation,
the corresponding BLAS or LAPACK routine and its associated cost.

\begin{center}
\renewcommand{\lstlistingname}{Algorithm}
\renewcommand{\algname}{}
\begin{minipage}{0.92\linewidth}
\renewcommand{\arraystretch}{1.2}
\centering
\large
\begin{lstlisting}[caption=\eig{},label=alg:core, escapechar=!]
$\mathcal{B} :=$ !{\sc \eig{}}!( $X_L$, $\mathcal{X}_R$, $\mathcal{Y}$, $h^2_j$, $\sigma^2_j$, $\Phi$ ) 
   $Z \Lambda Z^T = \Phi$                             !{\sc (eigendec)}!      $\frac{10}{3}n^3$
   $X'_L := Z^T X_L$                             !{\sc (gemm)}!         $2 n^2 (p-1)$
   $\mathcal{X}_R' := Z^T \mathcal{X}_R$                             !{\sc (gemm)}!         $2 n^2 m$
   $\mathcal{Y}' := Z^T \mathcal{Y}$                             !{\sc (gemm)}!         $2 n^2 t$
   for 1 $\le$ j $\le$ t
     $D_j := \sigma^2_j (h^2_j \Lambda + (1 - h^2_j) I)$                           !{(scalar-op)}!       $2n$
     $K_j K_j^T = D_j^{-\frac{1}{2}}$                           !{(scalar-op)}!       $2n$
     $W_{L_j} := K_j^T X'_L$                           !{(scalar-op)}!       $(p-1)n$
     $v_j := K_j^T y_j'$                           !{(scalar-op)}!       $n$
     $S_{TL_j} := W_{L_j}^T W_{L_j}$                           !{\sc (syrk)}!         $(p-1)^2 n$
     $b_{T_j} := W_{L_j}^T v_j$                           !{\sc (gemv)}!         $2 (p-1) n$
     for 1 $\le$ i $\le$ m
       $W_{R_{ij}} := K_j^T X'_{R_i}$                         !{(scalar-op)}!       $n$
       $S_{BL_{ij}} := W_{R_{ij}}^T W_{L_j}$                         !{\sc (gemv)}!         $2 (p-1) n$
       $S_{BR_{ij}} := W_{R_{ij}}^T W_{R_{ij}}$                         !{\sc (dot)}!          $2 n$
       $b_{B_{ij}} := W_{R_{ij}}^T v_j$                         !{\sc (dot)}!          $2 n$
       $b_{ij} := S_{ij}^{-1} b_{ij}$                         !{\sc (posv)}!         $O(p^3)$
\end{lstlisting}
\end{minipage}
\end{center}

\subsection{Computational cost}

Let us now compare the computational cost of \eig{} with 
that of the aforementioned alternatives:
LAPACK, \fastlmm{}, and \genabel{}. To this end, 
we recall the size of the input and output operands:

\begin{minipage}{0.3\textwidth}
\begin{itemize}
	\item $X_L \in R^{n \times (p-1)}$,
	\item $\mathcal{X}_R \in R^{n \times m}$,
\end{itemize}
\end{minipage}
\begin{minipage}{0.3\textwidth}
\begin{itemize}
	\item $\mathcal{Y} \in R^{n \times t}$,
	\item $h^2_j$, $\sigma^2_j \in R$,
\end{itemize}
\end{minipage}
\begin{minipage}{0.3\textwidth}
\begin{itemize}
	\item $\Phi \in R^{n \times n}$,
	\item $\mathcal{B} \in R^{m \times t \times p}$.
\end{itemize}
\end{minipage}

\noindent
Typical values for these dimensions are: 

\begin{minipage}{0.3\textwidth}
\begin{itemize}
	\item $10^3 \le n \le 10^4$,
	\item $2 \le p \le 20$,
\end{itemize}
\end{minipage}
\begin{minipage}{0.3\textwidth}
\begin{itemize}
	\item $10^6 \le m \le 10^8$, 
	\item $10^4 \le t \le 10^5$.
\end{itemize}
\end{minipage}

The asymptotical cost of \eig{} (see third column in Algorithm~\ref{alg:core}) is
$O(n^3 + m n^2 + t n^2 + t m p n)$. 
Since in a typical scenario for multi-trait GWAS $m$
and $t$ are much larger than $n$, the dominating factor is $O(t m p n)$. 
By contrast, the cost for the traditional library approach of LAPACK, 
which optimizes for a single GLS ($O(n^3)$)
and uses it for each point in the two-dimensional grid, is $O(t m n^3)$; 
the cost for state-of-the-art tools ---\fastlmm{} and \genabel{}---, which optimize for a one-dimensional analysis
($O(m n^2)$) and use it for each column in Fig.~\ref{fig:ProblemDescription}, is $O(t m n^2)$.

Table~\ref{tab:cost} collects the mentioned costs together with the
ratio with respect to \eig{}. 
The message is clear: No matter how
optimized a solver for a single GLS or for a one-dimensional analysis
is, it cannot compete with a solver specifically tailored for the
entire multi-trait analysis. 

\eig{} exploits the specific structure of
the operands and the correlation among GLSs, and
lowers the cost of
the best existing methods by a factor of $O(\frac{n}{p})$.
For a problem of size $n=1{,}000$, $p=4$ and large $m$ and $t$, we can expect \eig{} to be around two orders of magnitude faster 
than \flmm{} and \gw{}.

\begin{table*}
\renewcommand{\arraystretch}{1.4}
\centering
  \begin{tabular}{l@{\hspace*{8mm}} c@{\hspace*{8mm}} c} \toprule
	  & {Computational cost} & Ratio over \eig{} \\ \midrule
	  {\bf {\phantom{y}{\sc lapack}\phantom{y}} }   & $O(t m n^3)$ & $O(\frac{n^2}{p})$ \\[2mm]
	  {\bf {\phantom{y}{\sc gwfgls}\phantom{y}} }   & $O(t m n^2)$ & $O(\frac{n}{p})$   \\[2mm]
	  {\bf {\phantom{y}{\sc fast-lmm}\phantom{y}} } & $O(t m n^2)$ & $O(\frac{n}{p})$   \\[2mm]
	  {\bf {\phantom{y}\eig{} \phantom{y} } }       & $O(t m n p)$ & $1$  \\[2mm]
	\bottomrule
  \end{tabular}
\caption{Asymptotic cost of each of the discussed approaches to multi-trait GWAS.
The ratio over \eig{} illustrates the impact of exploiting increasing levels of correlation
within GWAS.
\eig{} improves the cost of state-of-the-art tools by a factor of $O(\frac{n}{p})$.}
\label{tab:cost}
\end{table*}

\section{Out-of-core: Analysis of computation and data transfers}
\label{sec:ooc}

In addition to the formidable computational complexity, GWAS poses a
second challenge: the management of large datasets.  In the prospective
scenario in which 
$m= 36{,}000{,}000$, $t= 300{,}000$, and $n = 10{,}000$,
the size of input and output data amounts to tens and hundreds 
of terabytes, respectively. Current analyses
already involve the processing and generation of few terabytes of data.

Obviously, present shared-memory architectures are not equipped with
such an amount of main memory; the size of the datasets thus becomes a
limiting factor.
In order to overcome this limitation, we turn our attention to
out-of-core
algorithms~\cite{Toledo:1999:SOA:327766.327789,springerlink:10.1007/BF00154340,Agullo_towardsa}.
The goal is to design an algorithm that makes an effective use of the
available input/output (I/O) mechanisms, to deal with data sets as
large as the available secondary storage, and to minimize the overhead
due to data transfers.

In the general case, the operands $\mathcal{X}_R$, $\mathcal{Y}$, and $\mathcal{B}$
are too large to fit in main memory and need to be streamed from disk to memory
and vice versa.
A naive approach, which too often
becomes the choice for actual implementations, is sketched
in Algorithm~\ref{alg:coreio}.  The algorithm loads one $y$ and one
$X_R$ and stores one $b$ at a time, resembling an unblocked algorithm.
A quick study of the computation and data transfers suffices to
understand the poor quality of the approach.
Let us assume the following problem sizes: $n=1{,}000$, $p=4$, $m=10^6$,
and $t=10^5$.
On the one hand, this data set requires the performance of approximately 1.1 petaflops; 
at a rate of 25 GFlops/sec 
(see Sections~\ref{sec:parallelism} and~\ref{sec:experiments} for details), 
it would complete in about 12 hours.
On the other hand, 
loading $t$ times the whole operand $\mathcal{X}_R$
($n \times m \times t \times 8$, assume 8 byte, double precision, data)
already causes 800 TBs of traffic 
between disk and main memory.
At a peak bandwidth of 2 GB/sec, which is by no means reached due to the small size of the transfers, 
the data movement alone takes 111 hours, 
about 10 times more than the time spent in actual computation. 
In order to attain an efficient out-of-core design, one has to 
a) study how different parameters affect the ratio between computation 
and data movement, aiming at reducing the latter, and 
b) use techniques for overlapping I/O with computation,
thus leading to a complete elimination of I/O overhead.
\begin{center}
\renewcommand{\lstlistingname}{Algorithm}
\renewcommand{\algname}{}
\begin{minipage}{0.92\linewidth}
\renewcommand{\arraystretch}{1.2}
\centering
\large
\begin{lstlisting}[caption=Sketch of a naive out-of-core scheme for \eig{}.,label=alg:coreio, escapechar=!]
$\mathcal{B} :=$ !{\sc \eig{}}!( $X_L$, $\mathcal{X}_R$, $\mathcal{Y}$, $h^2_j$, $\sigma^2_j$, $\Phi$ ) 
   !{\tt [$\ldots$]}!
   for 1 $\le$ j $\le$ t
     !{\tt load $y_j$}!
     !{\tt [$\ldots$]}!
     for 1 $\le$ i $\le$ m
       !{\tt load $X_{R_i}$}!
       !{\tt [$\ldots$]}!
       !{\tt store $b_{ij}$}!
\end{lstlisting}
\end{minipage}
\end{center}

The key behind a drastic reduction of data movement is data reuse;
this is commonly attained by blocked algorithms or, in the context of
out-of-core, {\em tiled} algorithms.  The idea
consists in loading not one single $y$ and $X_R$ at a time, but many of
them, in a chunk or {\em slab}; once loaded in memory, each of the
elements in the slabs can be used repeatedly.  Following this
approach, the cost of an expensive operation such as a disk-memory transfer,
is amortized by performing many more 
arithmetic operations per transferred element.

The ratio of computation over transferred data
\begin{equation}
r \equiv \frac{\text{\tt\# flops}}{\text{\tt data\_to\_load + data\_to\_store}}
\label{eq:ratio}
\end{equation}
gives an approximate idea of the potential for the minimization
of the impact of the data transfers.
Since the time for loading an element from disk is much larger than
performing a scalar operation, large ratios are desired.
When applied to a concrete situation, the ratio $r$ exposes a number of parameters
or {\em degrees of freedom} that may be adjusted to improve the ratio. In the context
of Algorithm~\ref{alg:core}, these degrees of freedom are the number of $y$'s and
$X_R$'s loaded at a time.

The above ratio may be extended to incorporate the concept of overlapping. 
For the algorithm to completely hide the I/O under 
computation, the time spent in computation must be larger than the time
for loading and storing data. Hence, the inequality
$$
\nonumber
\text{\tt computation\_time} > \text{\tt IO\_time}
$$
must hold. The inequality may be refined as
\begin{equation}
\frac{\text{\tt\# flops}}{\text{\tt\# flops/sec}} > \frac{\text{\tt data\_to\_load + data\_to\_store}}{\text{\tt IO\_bandwidth}}.
\label{eq:overlap}
\end{equation}
Inequality~\eqref{eq:overlap} enables the identification of (ranges of) values
for the aforementioned degrees of freedom, so that a perfect overlapping is achieved.

As an overlapping out-of-core 
mechanism we choose the so-called {\em double-buffering}. 
In short, to deploy double-buffering, main memory is divided into two
workspaces. Each workspace contains buffers, one
per operand to be streamed; in this case, two buffers for the input
operands $\mathcal{X}_R$ and $\mathcal{Y}$, and one for the output
stream $\mathcal{B}$. At a given iteration over the streams, one
workspace is used for computation, while the other one is used for
downloading previous results and uploading data for the next
iteration. After each iteration, the workspaces swap roles. For more
details, we refer the reader to~\cite{SingleGWAS}.

The rest of this section is dedicated to the analysis of the computation and data transfer required by
Algorithm~\ref{alg:core}. During the analysis we expose the 
degrees of freedom, and provide the specific constraints they must satisfy.

\subsection{Preloops: Single sweep over the input streams}
\label{subsec:preloop-ooc}

We logically divide Algorithm~\ref{alg:core} in two sections: 1) {\it preloops} ---lines 1 to 4---, and
2) {\it loops} ---lines 5 to 17---. In the preloops, the operations at lines 1 and 2 
involve operands that fit in main memory; 
they are therefore computed through direct calls
to the corresponding BLAS and LAPACK routines. 
On the contrary, the two matrix products at 
lines 3 and 4 involve the operands $\mathcal{X}_R$ and $\mathcal{Y}$, which reside on disk, 
and must therefore be performed in a streaming fashion.

The approach to compute $\mathcal{X}'_R := Z^T \mathcal{X}_R$ (line 4), 
and equivalently $\mathcal{Y}' := Z^T \mathcal{Y}$ (line 3),
consists in a traversal over the stream 
$\mathcal{X}_R$ in slabs containing $nb$ $X_R$'s, 
where the optimal value for $nb$ is to be estimated.
At every step over the stream (Fig.~\ref{fig:sweep}), the algorithm
\begin{enumerate}
\item loads $nb$ $X_R$'s, each of them of size $n$, i.e., $n \times nb$ elements;
\item stores $nb$ $X'_R$'s, each of them also of size $n$; and 
\item performs $2 \times n^2 \times nb$ flops, corresponding to the
  matrix product $\mathcal{X}'_{R_i} := Z^T \mathcal{X}_{R_i}$.
\end{enumerate}
The ratio of computation over data movement is
$$ \frac{2 \times n^2 \times nb}{(n \times nb) + (n \times nb)} \quad \equiv \quad n. $$
For typical values of $n$ ---from one thousand to tens of
thousands---, such a ratio shows great potential for perfect
overlapping.

Notice that, even though the ratio is independent of the value of $nb$, 
two quantities in~\eqref{eq:overlap} are influenced by this parameter: 
the performance of {\sc gemm} ({\tt \# flops/sec}), 
and the data transfer rate ({\tt IO\_bandwidth}).
The objective for both quantities is clear: 
One wants to maximize them both, 
to minimize execution time and  I/O time.
The constraint to satisfy is
\begin{equation}
	\frac{2 \times n^2 \times nb}{\text{\tt\# flops/sec}} > \frac{2\times n\times nb\times \text{\tt sizeof(datatype)}}{\text{\tt IO\_bandwidth}},
\end{equation}
which simplifies to
\begin{equation}
	\frac{n}{\text{\tt\# flops/sec}} > \frac{\text{\tt sizeof(datatype)}}{\text{\tt IO\_bandwidth}}.
\label{eq:nb}
\end{equation}
Since these quantities are specific to the architecture and the problem,
we defer their evaluation to Sec.~\ref{sec:experiments},
where the experimental setup is defined.

\subsection{Loops: Cross product of the input streams}

\begin{figure}
    \centering
	\begin{subfigure}{0.45\textwidth}
		\centering
		\vspace{6mm}
		\includegraphics[scale=.70]{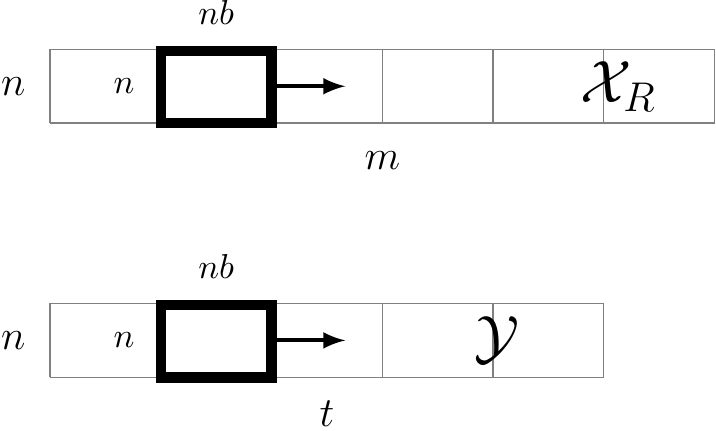}
		\caption{Single-sweep traversal of the streams $\mathcal{X}_R$ and $\mathcal{Y}$ in slabs of $nb$.}
		\label{fig:sweep}
	\end{subfigure}
	\hfill
	\begin{subfigure}{0.45\textwidth}
		\centering
		\includegraphics[scale=.65]{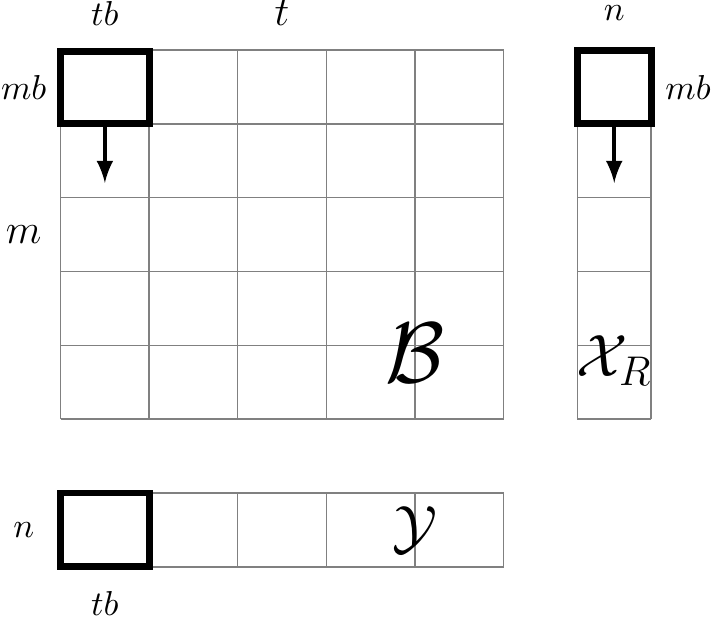}
		\caption{Computation of object $\mathcal{B}$ of size $m \times t$ decomposed into tiles of size $(mb \times tb)$. Traversal by columns.}
		\label{fig:tiling}
	\end{subfigure}
	\caption{Meaning of the degrees of freedom in the application of double-buffering to \eig{}.}
	\label{fig:degrees-freedom}
\end{figure}

The analysis and application of double-buffering to the second section of the algorithm requires a deeper discussion.
Instead of a single sweep through the streams, this section operates on every pair ($X_{R_i}$, $y_j$) of data,
i.e., a ``cross product'' of the input streams $\mathcal{X}_R$ and $\mathcal{Y}$. 
The problem can be translated into how the object $\mathcal{B}$ is built. 
In view of the impossibility of fitting the entire data set in memory, the problem of computing $\mathcal{B}$ is 
decomposed into the computation of smaller parts or tiles (Fig.~\ref{fig:tiling}).
Let us assign each tile the size $mb \times tb$, 
where $mb$ and $tb$ are the number of $X_R$'s and $y$'s processed, respectively. Henceforth we concentrate on determining adequate values for these two parameters.

As for the first section of the algorithm, the goal is to achieve a perfect overlapping
of I/O with computation. To this end, we must once more study the ratio of computation over 
data movement as a function of the tile size. 
To compute a tile, the algorithm must load $(mb \times n) + (tb \times n)$ elements, 
corresponding to the loading of a slab of $\mathcal{X}_R$'s and a slab of $\mathcal{Y}$'s,
and store $(mb \times tb \times p)$ elements, corresponding to a slab of $\mathcal{B}$. 
Per tile, $O(mb \times tb \times n \times p)$ flops are performed, 
for a ratio of 
\begin{equation}
	\nonumber
	\frac{O(mb \times tb \times p \times n)}{(mb \times n + tb \times n) + (mb \times tb \times p)}.
\label{eq:full-ratio}
\end{equation}
In Section~\ref{sec:tailoringtwod} we deduced that a traversal by columns of the grid
depicted in Fig.~\ref{fig:ProblemDescription}, is favorable in terms of computational
cost and temporary storage. Hence, the algorithm
will load a slab of $\mathcal{Y}$'s and reuse it
for all $\mathcal{X}_R$'s. Consequently, the cost of loading the slab of $\mathcal{Y}$'s can be neglected, and
the ratio simplifies to
\begin{equation}
	\nonumber
	r = \frac{O(tb \times p \times n)}{n + tb \times p}.
\label{eq:tb-ratio}
\end{equation}

\begin{figure}
    \centering
	\includegraphics[scale=.45]{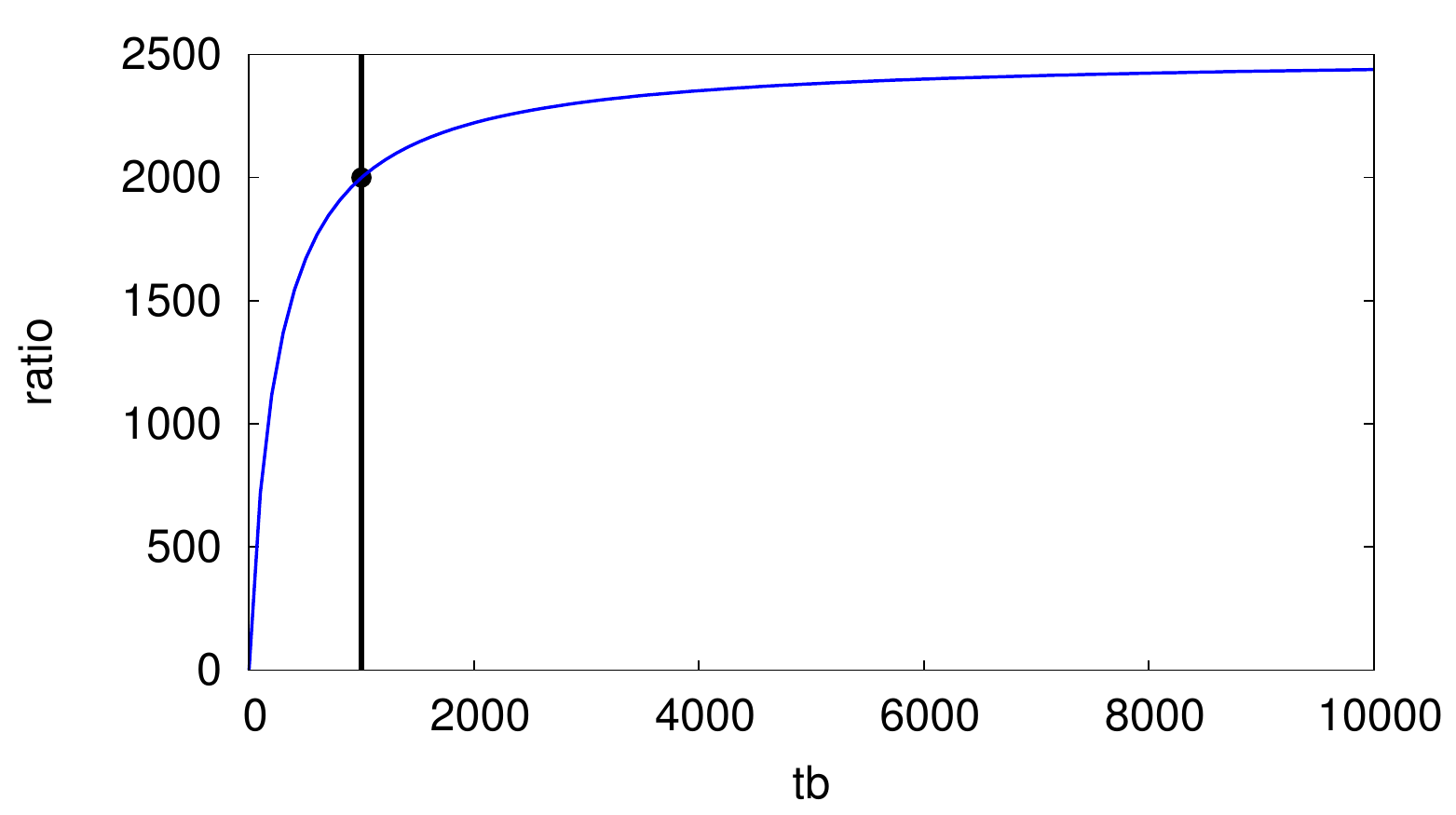}
	\caption{Ratio of computation over data transfer for tiles of $\mathcal{B}$.
	         The ratio is plotted as a function of $tb$, with $p=4$ and $n=1000$.
		     The ratio is initially very low but grows rapidly to values that allow
		     a perfect overlapping of I/O with computation.}
\label{fig:tb-ratio}
\end{figure}

This ratio $r$ is independent of the value $mb$. Intuitively, if $mb$ is multiplied by 2, 
both the amount of computation within the tile and the required I/O are doubled 
(twice the number of $X_R$'s are loaded and twice the number of $b_{ij}$'s are computed and stored).
On the contrary, the ratio grows monotonically with $tb$. 
For the sake of clarity, we illustrate in Fig.~\ref{fig:tb-ratio} the behavior of $r$
as a function of $tb$.
For $tb = 1$, $r$ is $O(p)$; 
the ratio rapidly grows with $tb$ until it reaches a value of about $2n$,
from where the growth is much smaller.
As discussed previously, large values of the ratio are favored.
Therefore, tiling along the $t$ dimension ($tb > 1$) is imperative to reduce the I/O overhead.

However, since we are making use of double-buffering, we do not need to choose
the largest possible $tb$;
we only require $tb$ to be large enough to completely hide the overhead due to I/O.
The minimum value of $tb$ leading to a perfect overlap can be
determined analytically via~\eqref{eq:overlap}. 
The time for computing a tile must be larger
than the time for loading a slab of 
$\mathcal{X}_R$ and storing the corresponding slab of $\mathcal{B}$: 
$$
\frac{mb \times tb \times (5 + 2 (p-1)) \times n}{\text{\tt\# flops/sec}} > \frac{(mb \times n + mb \times tb \times p) \times \text{\tt sizeof(datatype)}}{\text{\tt IO bandwidth}}.
$$
Dividing both sides by $mb$:
\begin{equation}
\frac{tb \times (5 + 2 (p-1)) \times n}{\text{\tt\# flops/sec}} > \frac{(n + tb \times p) \times \text{\tt sizeof(datatype)}}{\text{\tt IO bandwidth}}.
\label{eq:tb}
\end{equation}

In Eq.~\eqref{eq:tb}, the values for $n$ and $p$ are given as input; 
the datatype is known;
the hard-drive bandwidth is to be determined empirically; and
the performance attained in the computation of a tile is determined
when tailoring for the architecture, 
as discussed in Section~\ref{sec:parallelism}.
As we observed for $nb$, the parameter $mb$ does not influence the reduction
of data transfer, and it has to satisfy no explicit constraint.
Still, tiling along $m$ ($mb > 1$) is recommended for performance reasons, as
we demonstrate in Section~\ref{sec:parallelism}. 

Let us summarize the study performed in this section. We exposed and analyzed the degrees of freedom
in the tiling of our algorithm: $nb$, $mb$, and $tb$. The main message of the study is that $tb$ is
the key in the reduction of the data movement to make the computation feasible. Additionally, we discussed
the implication of the value of all three parameters in terms of performance, and stated the two key
constraints ---Eqs.~\eqref{eq:nb}~and~\eqref{eq:tb}--- to be satisfied in order to completely eliminate overhead
due to I/O operations.

\section{Tailoring for shared-memory parallelism}
\label{sec:parallelism}

Algorithm~\ref{alg:core} (described in Sections~\ref{sec:core-algorithm}
and~\ref{sec:ooc}) both has a lower computational complexity than all 
the current alternatives and eliminates the overhead due to I/O.
Yet, a careful implementation is required to exploit shared-memory parallelism.
In this section, 
we explain how to select the best-suited type of parallelism 
for each section of the algorithm,
and outline the steps to achieve almost optimal scalability. 

\subsection{Preloop: Multi-threaded BLAS}

The operations in the preloop section, lines 2--5, correspond to 
either LAPACK routines (eigendecomposition) that cast most of 
the computation in terms of BLAS-3 kernels, or are direct calls to 
BLAS-3 routines ({\sc gemm}).
For this class of routines,
it is well known that optimized multi-threaded BLAS libraries 
deliver both high-performance and scalability. 
This is therefore the solution we adopt. 

\subsection{Loops: Single-threaded BLAS + OpenMP parallelism}

In sharp contrast to the preloops, the computation performed within
the loops, lines 6--18, maps to non-scalable BLAS-1 and
BLAS-2 operations, thus making the use of multi-threaded BLAS 
not viable. Instead, we exploit the multi-core parallelism by utilizing
a single-threaded BLAS in combination with OpenMP threads and 
by decomposing the computation in tasks.

In Section~\ref{sec:ooc}, it was shown that tiling 
the computation of $\mathcal{B}$
along the $t$ dimension is the key to eliminate penalties 
due to I/O data transfers. In this section we demonstrate 
that for a high-performance shared-memory implementation, 
it is also crucial to tile  along the $m$ dimension, 
and discuss how to select the best tile size and shape.
Furthermore, we explore different multi-threading strategies to manage
the data transfers, and to split and assign computational tasks to
cores; since the total number of options is daunting, here we
only describe those two that we considered most promising:
\begin{enumerate}
\item a single {\it master} thread is responsible for the
  streaming of all tiles, 
  and all threads collaborate in the computation of each tile; and
\item each thread operates on whole tiles, and is responsible for their
  streaming. 
\end{enumerate}

\subsubsection{Two-dimensional tiling}

If we only consider tiles of size $1 \times tb$, 
the largest tile is of size $1 \times 100{,}000$ 
(See Fig.~\ref{fig:tiling}, with $mb = 1$ and $tb = t$). 
In our experimental environment with 40 threads, each thread computes a
task or {\em block} of size $1 \times 2{,}500$, attaining poor
efficiency: 1.54\%
Instead, by tiling (and blocking) along $m$ as well as along $t$,
efficiency increases:
\begin{enumerate}
	\item [a)] For a tile size of $200 \times 100{,}000$ with blocks of size $1 \times 100{,}000$,
		the attained efficiency is 3.75\%.
	\item [b)] For a tile size of $1{,}000 \times 100{,}000$ with blocks of size $100 \times 100$,
		the efficiency raises to 6.35\%.
\end{enumerate}
A higher workload per thread ---case a)--- results in a speedup of
about 2.5x; by further increasing the value of $mb$ ---case b)---
data locality improves, for speedups of about 4x.
The benefits of using two-dimensional tiles are thus clear; 
it remains to determine how such tiles should be decomposed 
to maximize performance.

\subsubsection{Estimating the best block size}
Typically, as long as the tile size satisfies the constraints~\eqref{eq:nb} 
and~\eqref{eq:tb-ratio}, 
one chooses $mb$ and $tb$ to maximize the RAM usage. 
For performance purposes, the computation 
of such large tiles must be split into smaller 
blocks to 
exploit both the memory hierarchy and all the available cores.
The same way the tile size is chosen according to the amount of available
main memory, in cache-based architectures the block size is determined
according to the size of the cache memories. 
The challenge lies on finding the optimal block size $(mbb \times tbb)$.

\begin{figure}
  \centering
  \includegraphics[scale=.55]{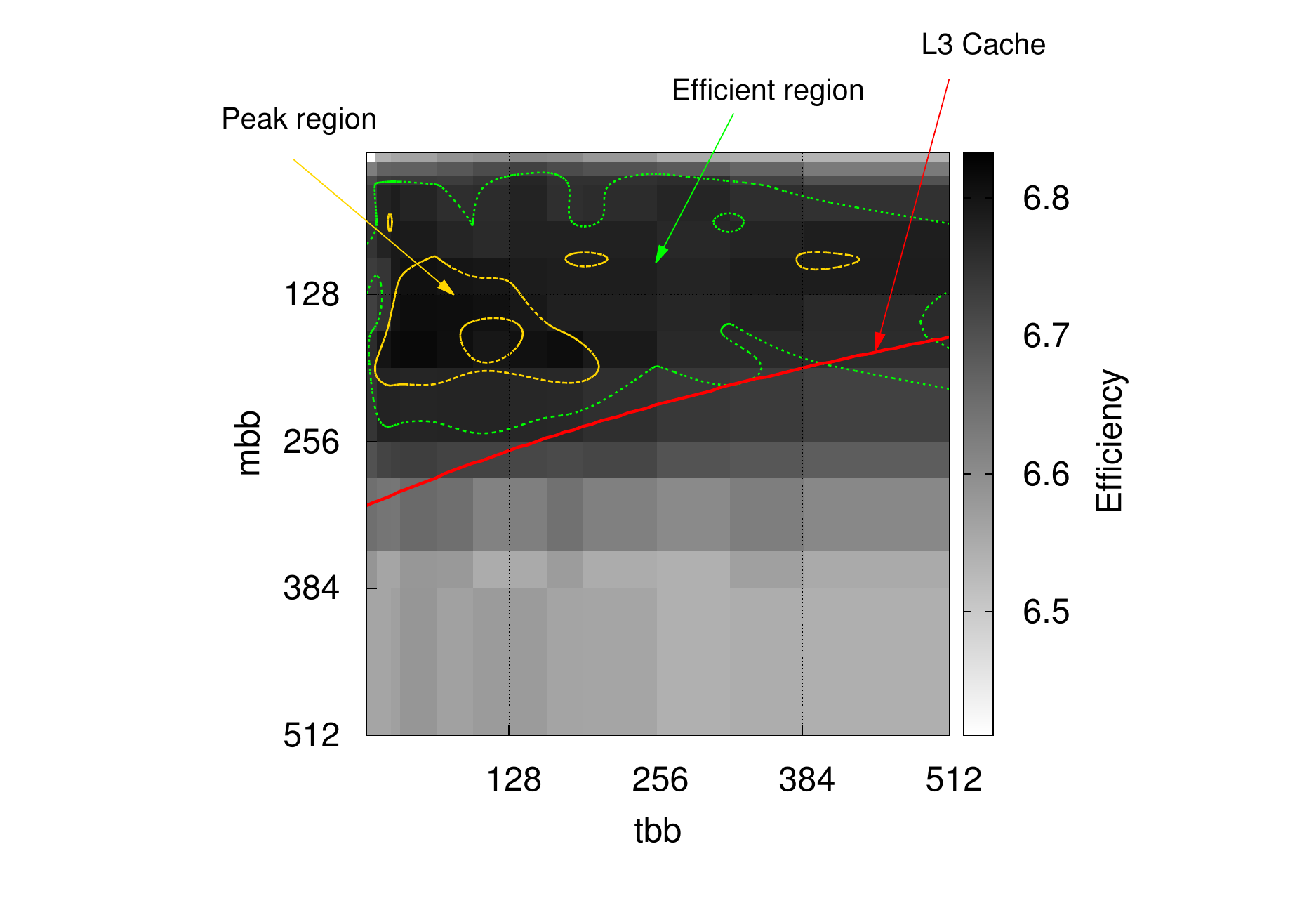}
  \caption{Study of the performance attained by different block
    sizes $(mbb \times tbb)$.
	The red line (L3 cache) delimits the blocks that fit in
    L3 cache ---above the line---. The green region
    (Efficient region) includes block sizes attaining an efficiency
	of at least 6.7\%. Most of the region fits in L3. The
    yellow region (Peak region) includes the most efficient blocks
    (efficiency above 6.8\%). The fluctuation within the Peak region
    is less than 0.5\%.}
\label{fig:blocksize}
\end{figure}

We consider blocking for the highest level of cache.
Fig.~\ref{fig:blocksize} provides a heatmap representing the
efficiency attained varying the block sizes.  The red line
delimits the space of block sizes that fit in the last level of
cache ---level 3 (L3) for the architecture used in our experiments
(see Section~\ref{sec:experiments} for details)---.  As anticipated,
the block sizes attaining higher efficiency (green line: {\it Efficient} region)
lie above the red line, i.e., fit in L3 cache. Also expected
is the fact that relatively large square blocks attain the highest
efficiency (yellow line: {\it Peak} region).

Interestingly, 
we observed that inside the Peak region performance plateaus, 
exhibiting variations below 0.5\%.  
This suggests that one
should focus on finding this region
rather than {\it the} most efficient block size. 
As a matter of fact, since within the Peak region 
the fluctuations due to the hardware and the operating system are 
greater than the differences in performance, 
it can be argued that a best block size does
not even exist.

\subsubsection{Work distribution}
\label{sec:workdist}

We conclude describing two possibly strategies to distribute the 
blocks among cores.  

\begin{figure}
    \centering
	\begin{minipage}[t]{0.45\textwidth}
		\centering
		\includegraphics[scale=.50]{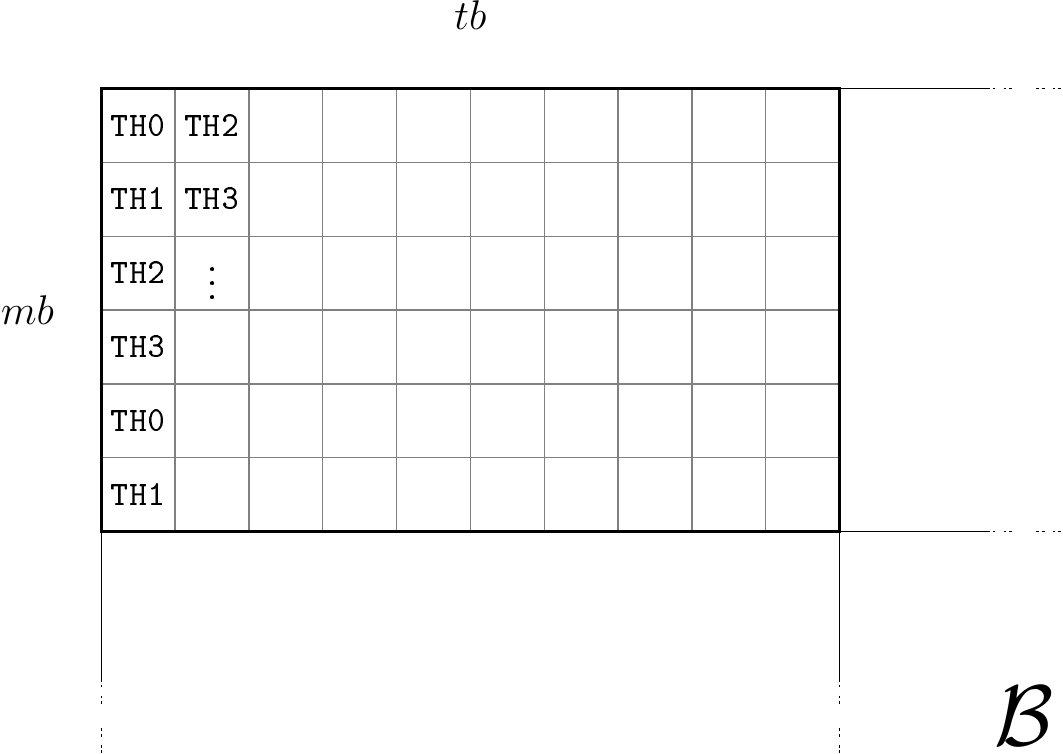}
		\subcaption{Cooperative threads: A master thread loads each tile, and all spawned threads cooperate on
		the computation of the tile.}
		\label{fig:one-tile}
	\end{minipage}
	\hfill
	\begin{minipage}[t]{0.45\textwidth}
		\centering
                \includegraphics[scale=.50]{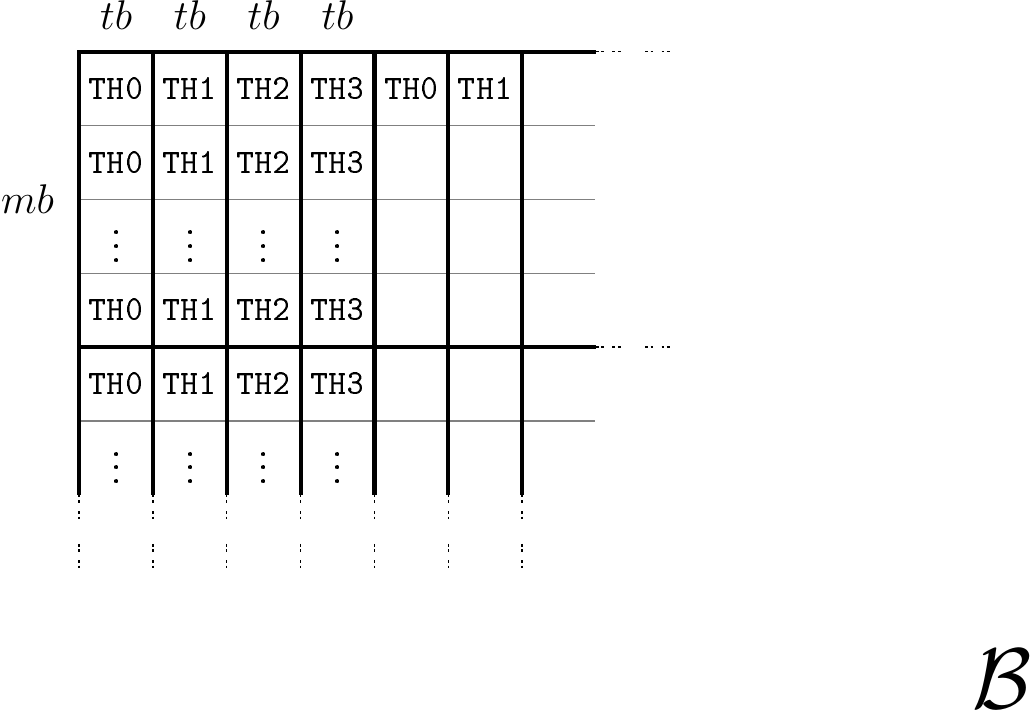}
                \subcaption{Independent threads: Each thread is
                  responsible for the loading, computation and
                  downloading of its own tiles.
                }
		\label{fig:multiple-tiles}
	\end{minipage}
	\caption{The two studied approaches to work distribution.}
\end{figure}

\begin{enumerate}
	\item {\it Cooperative threads} (\eigone{}).
	A common scheme for tile-based implementations of out-of-core algorithms
	consists of a master thread that takes care of all data transfers, and 
	a number of spawned threads that cooperate in the computation of each tile.
	As Fig.~\ref{fig:one-tile} illustrates, each tile is divided into blocks,
	and these are distributed among computing threads, thus sharing the workload. 

	\item {\it Independent threads} (\eigtwo{}). 
	Alternatively, work may be distributed so that each thread is responsible for
	the loading, computation, and downloading of its own entire tiles.
	As Fig.~\ref{fig:multiple-tiles} illustrates, each tile is still decomposed into blocks for performance reasons. 
\end{enumerate}

\section{Experimental results}
\label{sec:experiments}

We focus now on experimental results. We compare the performance and scalability
of the state-of-the art tools, \flmm{} and \gw{}, with those of the presented
algorithm \eig{}. For \eig{} we provide results for both parallel approaches
described in Section~\ref{sec:workdist}: \eigone{} for cooperative threads,
and \eigtwo{} for independent threads.

\subsection{Experimental setup}
All tests were run on a SMP system consisting of 8 Intel Xeon E7-4850
multi-core processors.  Each processor comprises ten cores, operating
at a frequency of 2.00~GHz, for a combined peak performance of
320~GFlops/sec.  The system is equipped with 512GB of RAM and 4TBs of
disk as secondary memory.  The I/O system attains a maximum bandwidth
of 2GBs/sec for data transfers of at least 2MBs.

The routines were compiled with the GNU C Compiler (gcc, version
4.4.5), and linked to a multi-threaded Intel's MKL library (version
10.3).  Our routines make use of the OpenMP parallelism provided by
the compiler through a number of {\it pragma} directives. All
computations were performed in double precision.

\subsection{Configuring the degrees of freedom}

In order to attain maximum performance, 
we need to estimate the most effective values 
for the algorithm parameters:
$nb$, $mb$, $tb$, $mbb$, and $tbb$. 

\subsubsection{$nb$}

As described in Section~\ref{sec:ooc}, this parameter is chosen to be
greater than or equal to the minimum value that maximizes both data
transfer rate and {\sc gemm}'s performance.  The maximum I/O bandwidth
is attained for transfers of at least 2MBs; the minimum value of $nb$
to reach this size is 250 ($nb \times n \times \text{\tt
  sizeof(datatype)} \equiv 250 \times 1000 \times 8B \equiv 2$MBs).
We also determined empirically that the minimum value of $nb$ to
maximize {\sc gemm}'s performance for 40 cores in the specified
architecture is $10{,}000$ (attaining 240 GFlops/sec). Therefore, we
set $nb$ to $10{,}000$.  Substituting in Eq.~\eqref{eq:nb}, we see
that we will achieve a perfect overlapping:
$$\frac{1000}{240 \times 10^9} > \frac{8}{2 \times 10^9} \equiv 4.17 > 4.$$

\subsubsection{$mbb$ and $tbb$}

According to the results shown in Fig.~\ref{fig:blocksize}, we simply choose a 
square block size within the Peak region: $160 \times 160$.

\subsubsection{$mb$ and $tb$}

We determined that the maximum performance attained in the computation of a tile is about
25 GFlops/sec. Substituting each variable in Eq.~\eqref{eq:tb}:
$$
	\frac{tb \times 11 \times 1000}{25 \times 10^9} > \frac{8000 + tb \times 32}{2 \times 10^9} \quad \equiv \quad tb > 10.
$$
Therefore, and given the chosen block size, the constraint above ($tb > 10$) does not impose any restriction
in our choice of tile size ($tb$ will be at least 160). 
In any case, we emphasize the importance of always carrying out such an analysis, as
differences in performance or I/O bandwidth could lead to more restrictive constraints.

As discussed in Section~\ref{sec:parallelism}, we choose different
tile sizes for the different approaches to work distribution. In the
case of \eigone{}, we consider large square tiles maximizing the usage
of free main memory; accordingly we choose $mb = tb = 25{,}600$.  For
\eigtwo{}, we fix the value of $tb$ to that of the block size, i.e.,
$tb = tbb = 160$, while for $mb$ we select a small multiple of the
block size ($mb=1600$). The reason for such a small tile size is to
show that our routine can achieve impressive speedups even with a
limited amount of main memory, emphasizing that the only restriction
for \eig{} is the size available amount of secondary device storage.
As an example, for the execution of the largest test reported in this section, 
which involves the processing of more than 3 terabytes of data, \eigtwo{} only required
less than 2 gigabytes of memory.

\subsection{Experiments}

We present performance results for \flmm{}, \gw{}, \eigone{}, and \eigtwo{}. 
At first, results for a single core are given, to highlight
the speedup due to the sole improvement of the algorithm, 
i.e., the reduction in computational cost. Then,
we compare the scalability and performance of all four algorithms, 
using all the available 40 cores.

\begin{figure}
    \centering
	\includegraphics[scale=.5]{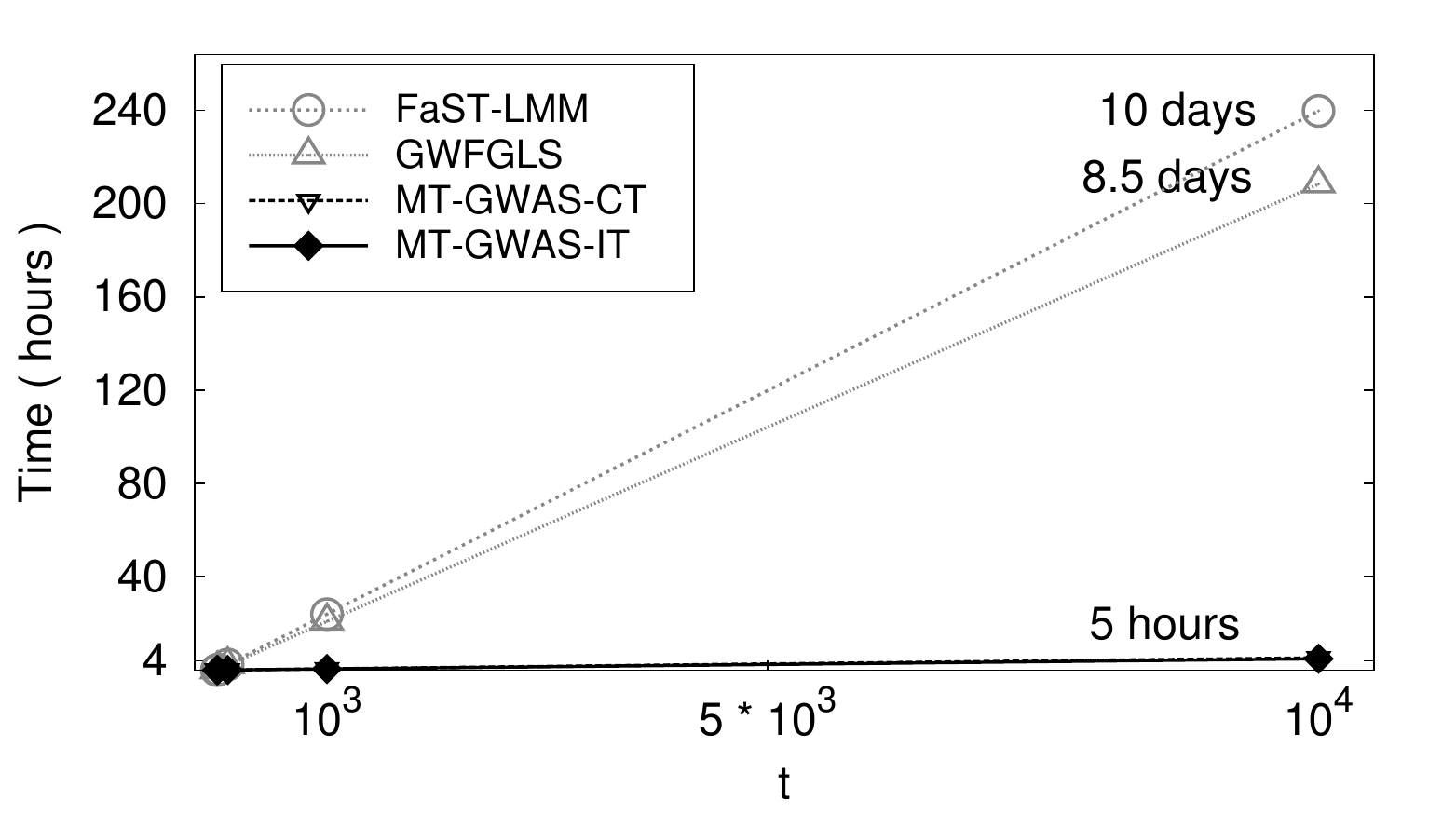}
	\caption{Performance of the single-threaded versions of the four presented routines. 
		The problem dimensions are: $n=1{,}000$, $p=4$, and $m=100{,}000$.
		For $t=10{,}000$, our routines ---\eigone{} and \eigtwo{}--- outperform the 
	state-of-the-art tools \flmm{} and \gw{} by a factor of 50 and 43, respectively.}
	\label{fig:single-core}
\end{figure}

In Fig.~\ref{fig:single-core}, we provide timings for the four routines 
for increasing values of $t$. 
The experiments were run using a single thread.
The gap between the state-of-the-art routines and our novel algorithm is substantial: 
while \flmm{} and \gw{} would take, respectively, about 10 and 8.5 days to complete,
\eigone{} and \eigtwo{} reduce the execution time to 5.3 and 4.8 hours, respectively.
In terms of speedups, our best routine, \eigtwo{}, is 50 and 43 times faster than \flmm{}
and \gw{}, respectively. It is also worth noticing
the 10\% speedup of \eigtwo{} with respect to \eigone{}.

\begin{figure}
    \centering
	\includegraphics[scale=.5]{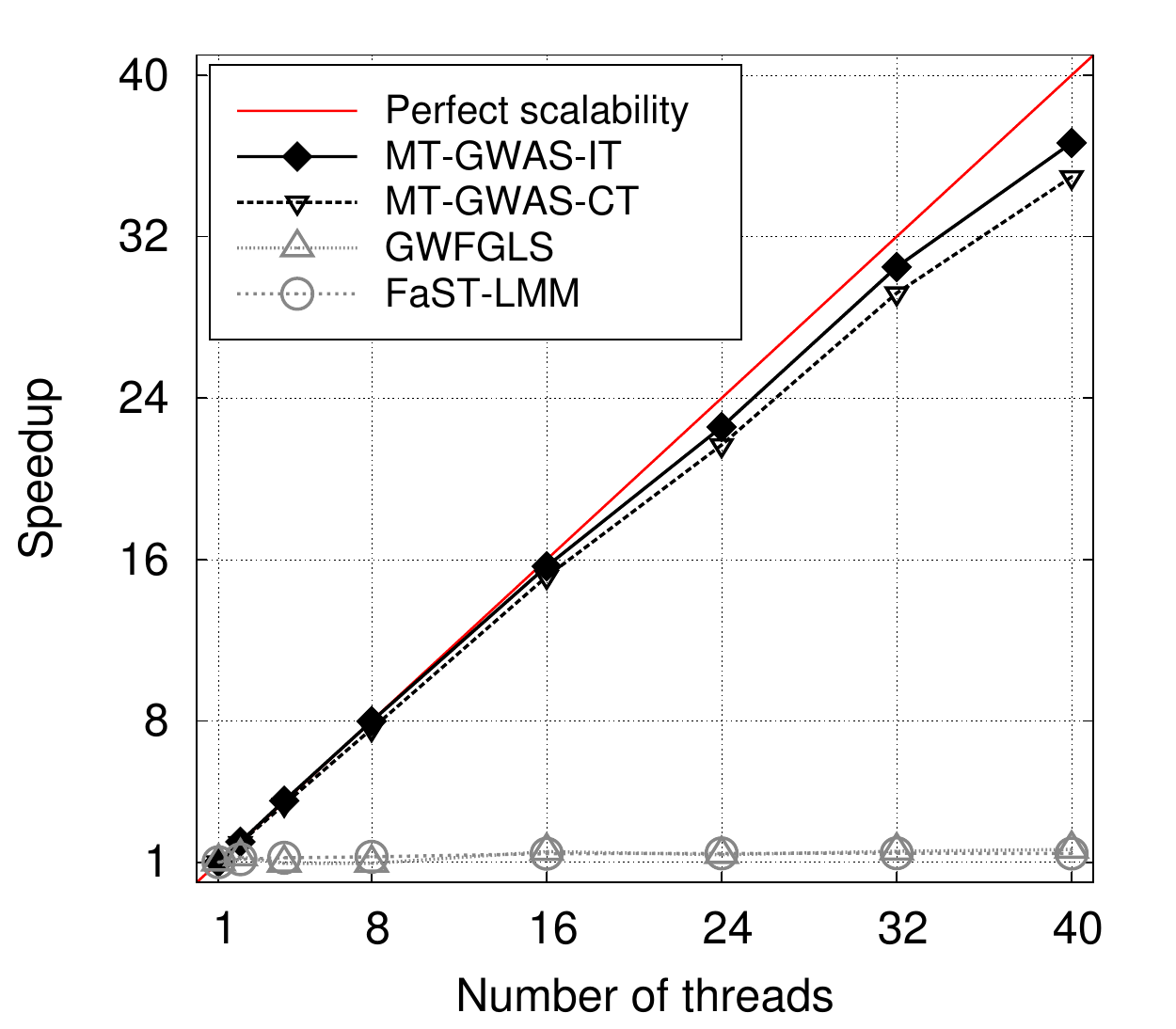}
	\caption{Scalability of \eigone{}, \eigtwo{}, \flmm{}, and \gw{}. 
		While \flmm{} and \gw{} present a mediocre scalability, 
		our routines ---\eigone{} and \eigtwo{}--- attain speedups of
		35x and 36.6x, respectively, when using 40 threads.
	The problem dimensions are: $n=1{,}000$, $p=4$, $m=100{,}000$, and $t=20{,}000$.} 
	\label{fig:scal}
\end{figure}

Scalability results are shown in Fig.~\ref{fig:scal}. 
The figure reveals clear scalability
issues in the parallel implementations of both \flmm{} and \gw{}; they achieve speedups
between 1.4 and 1.6, and plateau when 16 or more cores are used. 
Instead, the results for our routines clearly demonstrate the benefit of carefully tailoring Algorithm~\ref{alg:core} 
for large shared-memory architectures:
Using all available 40 cores, \eigone{} and \eigtwo{} attain speedups of 35 and 36.6,
respectively.
Furthermore, the trend presented by both our routines forecasts larger speedups, 
should more cores be available.

\begin{figure}
    \centering
	\includegraphics[scale=.5]{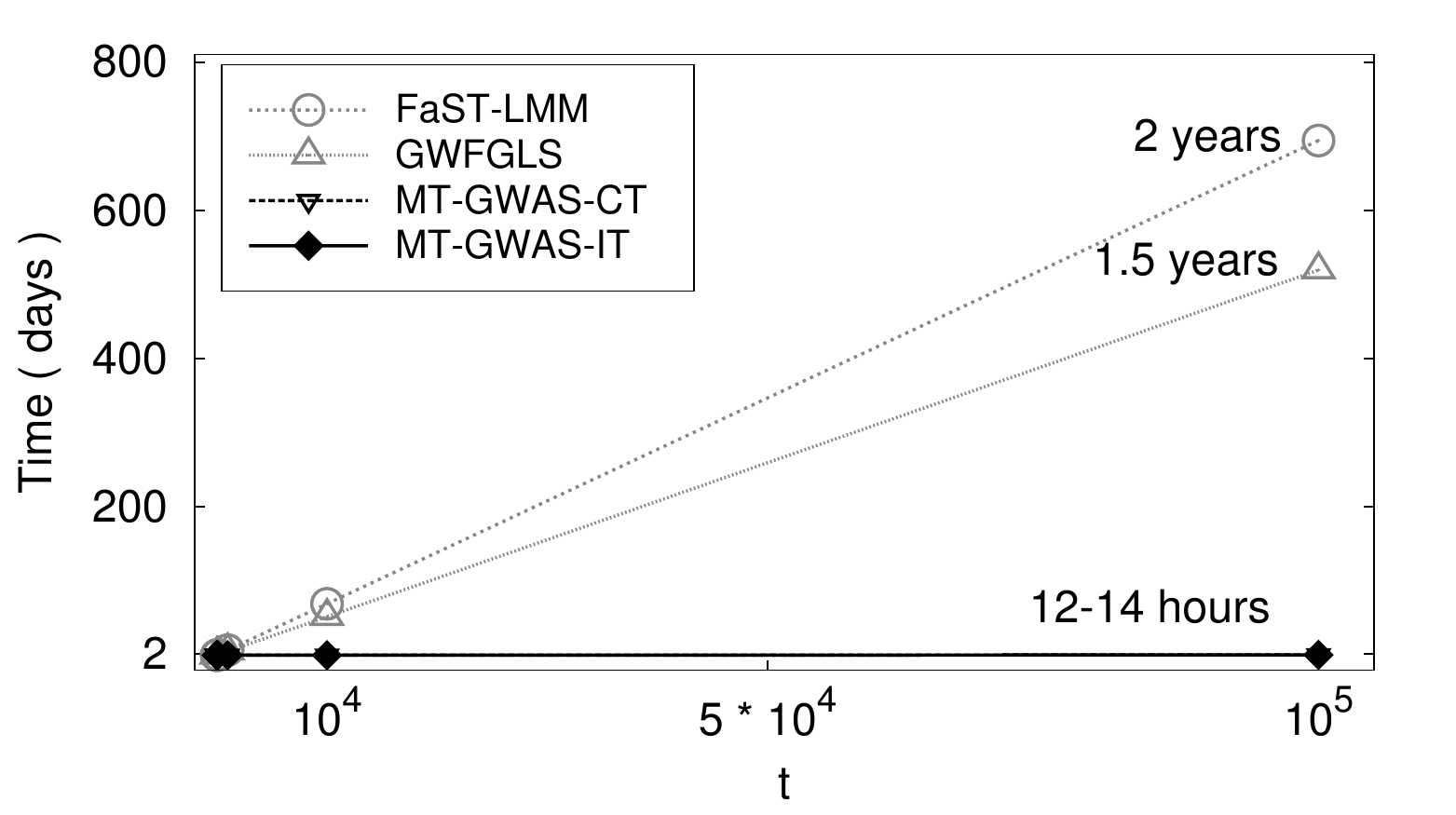}
	\caption{Performance of the multi-threaded versions.
		The problem dimensions are: $n=1{,}000$, $p=4$, and $m=1{,}000{,}000$.
		While, for $t=100{,}000$, \flmm{} and \gw{} are not viable, our routines
		complete in a matter of hours. The observed speedups are larger than 1000x.}
	\label{fig:multi-core}
\end{figure}

Figure~\ref{fig:multi-core} presents performance results for the same four routines 
when using 40 threads. 
Here, the effects of the computational cost reduction, the perfect
I/O overlapping, and a better scalability are all combined, yielding 
speedups of $1{,}352$ and $1{,}012$ over \flmm{} and \gw{}, respectively.
The time to compute the largest presented problem,
$n= 1{,}000$, $p=4$, $m=10^6$, and $t=10^5$, is reduced from (unfeasible) years
to 12 hours.

\section{Future Work}
\label{sec:futurework}

Multi-trait GWAS is constrained by three dimensions:
$n$, ranging from $10^3$ to $10^4$, 
$m$, ranging from $10^6$ to $10^8$, and
$t$, ranging from $10^4$ to $10^5$.
The work presented in this paper allows $m$ and $t$ to grow as large as desired.
On the contrary, our routines assume the operand $\Phi \in R^{n \times n}$ to fit in 
main memory, and thus are constrained by the size of $n$. For very large values of 
$n$, the problem demands a distributed-memory version of \eig{}, and therefore 
requires an extension of the analysis of data transfers and work distribution undergone in this paper.

Additionally, there is an increasing demand for support of co-processors 
such as GPUs. While the use of GPUs was proven successful for the single-trait 
case ($t = 1$)~\cite{Beyer2012:400},
routines for the more general multi-trait case are not yet available.
There, the challenge lies in writing  optimized kernels
for the computation within the loops, tuning for the intricate memory hierarchy of
the architecture.

\section{Conclusions}
\label{sec:conclusions}

We addressed an extremely challenging and widespread problem in computational
biology, the genome-wide association study (GWAS) with multiple
traits. GWAS involves large-scale computations ---petaflops--- on 
large data sets ---terabytes of data---, and the existing
state-of-the-art tools are only effective in conjunction with
supercomputers. In this paper we introduced \eig{}, a novel algorithm 
for sequences of least-squares problems,
tailored to take advantage of both application-specific knowledge 
and shared memory parallelism, 
and demonstrated that for performing the full GWAS analysis, 
a single multi-core node suffices. 

First, we described the derivation of an algorithm that exploits all knowledge 
available from the application: from the specific two-dimensional sequence of 
generalized least-squares problems, to the special structure of the operands. 
By eliminating redundant computations, this algorithm lowers the asymptotical 
cost of state-of-the-art tools by several orders of magnitude.

Then, we discussed how to deal with large-scale datasets. In order to 
incorporate an out-of-core mechanism, we analyzed the ratio between data 
movement and computations, and derived the best tile size and shape
for a perfect overlapping of data transfers with computation. This mechanism
enables the processing of data sets as large as the available secondary storage,
without any overhead due to I/O operations.

Finally, we tailored our algorithm for shared-memory parallel
architectures. The study of the different sections of the algorithm
suggested the use of a mixed parallelism: 1) multi-threaded BLAS, and
2) single-threaded BLAS and OpenMP parallelism. We empirically
estimated the best size for the computational tasks, and studied 
two different approaches to distribute those task among threads. 
The resulting routines attain speedups of 35x and 36.6x on 40 cores.

By combining the effects of the computational cost reduction, the perfect
I/O overlapping, and a high scalability, our routines yield, when compared to the
state-of-the-art tools, a 1000-fold reduction in the time to solution.
Thanks to this algorithm, analyses that were thought to be feasible 
only with the help of supercomputers,
can now be completed in matter of a few hours with a single multi-core node.

\section{Acknowledgements}

Financial support from the 
Deutsche Forschungsgemeinschaft (German Research Association) through
grant GSC 111 is gratefully acknowledged. 
The authors thank Matthias Petschow for discussion on the algorithms, and
the Center for Computing and Communication at RWTH Aachen for the computing resources.


\bibliography{bibliography}

\bibliographystyle{acmtrans}

\begin{received}
Received Month Year;
revised Month Year;
accepted Month Year
\end{received}
\end{document}